\documentstyle[preprint,aps,amsfonts]{revtex}

\begin{document}
\draft
\tightenlines

\preprint{\vbox{\hbox{WIS-94/2/Jan-PH}
                \hbox{hep-ph/9401246}
                \hbox{January 1994} }}

\title{Phenomenological Constraints on $\bar\Lambda$ and $\lambda_1$}

\author{Zoltan Ligeti and Yosef Nir}

\address{Department of Particle Physics \\
         Weizmann Institute of Science \\
         Rehovot 76100, Israel}

\maketitle

\begin{abstract}
Combining the experimental data on the inclusive decays
$D\to X\,e\,\nu$, $B\to X\,e\,\nu$ and $B\to X\,\tau\,\nu$, we find
severe constraints on the $\bar\Lambda$ and $\lambda_1$ parameters
of the Heavy Quark Effective Theory. In particular, we get
$\bar\Lambda<0.7\,$GeV. Our constraints further imply
$m_c\geq1.43\,$GeV and $m_b\geq4.66\,$GeV. We discuss future prospects
of improving these bounds and their phenomenological implications.
\end{abstract}

\newpage
\narrowtext

The Heavy Quark Effective Theory (HQET) provides a systematic way
of calculating $1/m_Q$ corrections to the model independent results
of heavy quark symmetry \cite{review}.  For inclusive semileptonic
$B$ and $D$ decay rates, Chay {\it et al.} \cite{CGG} showed that the
heavy quark limit ($m_Q\to\infty$) coincides with the free quark decay
model and that there are no corrections to this result at order $1/m_Q$.
The $1/m_Q^2$ corrections can be computed in terms of only two
additional nonperturbative parameters \cite{incl}.

HQET provides much more precise predictions for inclusive
semileptonic decays than the spectator quark model \cite{us}.
One ingredient in the improvement is the incorporation of the
nonperturbative $1/m_Q^2$ corrections which are, however,
numerically quite small. More important is that HQET gives an
unambiguous definition of the quark masses \cite{FNL}, and these
masses determine the inclusive decay rates.
Using the relation between the masses of heavy quarks in terms of the
HQET parameters $\bar\Lambda$, $\lambda_1$ and $\lambda_2$ (see their
definitions below) leads to a significant reduction in the theoretical
uncertainties \cite{us}.

These HQET parameters are genuinely nonperturbative: at present they
can only be estimated in models of QCD.  In ref. \cite{us} we used
such theoretical ranges for the various parameters to calculate
inclusive semileptonic $B$ decay rates.
Here we take the opposite approach and investigate what restrictions
can be derived from the existing experimental data on these parameters.
Our constraints help to pin down the physical values of the HQET
parameters. This will allow testing various model calculations and will
lead to significantly reduced theoretical errors in the predictions for
inclusive heavy hadron decays.

We now turn to an explicit presentation of the relevant HQET parameters.
In the framework of HQET, semileptonic decay rates depend on three
parameters (besides the Fermi constant, CKM angles, and the well-known
masses of physical particles): $\bar\Lambda$, $\lambda_1$, and
$\lambda_2$. $\bar\Lambda$ is given by the matrix element \cite{FNL}
\begin{equation}\label{Ldef}
\bar\Lambda = {\langle0| \bar q\,iv\!\cdot\!\overleftarrow{D}\,\Gamma\,
h_v |M(v)\rangle \over \langle0|\bar q\,\Gamma\,h_v|M(v)\rangle}\,,
\end{equation}
where $M(v)$ denotes the meson state and $h_v$ denotes the quark
field of the effective theory with velocity $v$.  This parameter
describes, at leading order in $1/m_Q$, the mass difference between
a heavy meson and the heavy quark that it contains, and sets the
scale of the $1/m_Q$ expansion.  The dimensionful constants
$\lambda_1$ and $\lambda_2$ \cite{FaNe} parameterize the matrix
elements of the kinetic and chromomagnetic operators, respectively,
which appear in the Lagrangian of HQET at order $1/m_Q$:
\begin{mathletters}\label{ldef}
\begin{eqnarray}
\lambda_1 &=& {1\over2}\langle M(v)|\bar h_v(iD)^2 h_v|M(v)\rangle\,,\\
\lambda_2 &=& {1\over2d_M}\langle M(v)|{g_s\over2}\bar h_v
\sigma_{\alpha\beta} G^{\alpha\beta} h_v|M(v)\rangle\,,
\end{eqnarray}
\end{mathletters}%
where $d_P=3$ and $d_V=-1$ for pseudoscalar and vector mesons,
respectively. Whereas $\lambda_1$ is not renormalized due to
reparameterization invariance \cite{LuMa}, $\lambda_2$ depends
on the scale.

The $d_M$ factor appears because the chromomagnetic operator breaks
the heavy quark spin symmetry.  Consequently, the value of $\lambda_2$
can be easily extracted from the mass splitting between the vector and
pseudoscalar mesons:
\begin{equation}
\lambda_2 ={m_{B^*}^2-m_B^2\over4} \simeq 0.12\,{\rm GeV}^2\,.
\end{equation}
We expect this value of $\lambda_2$ to be accurate to within 10\%, as a
result of the finite $b$ quark mass and the experimental uncertainties.
There is no similarly simple way to determine $\lambda_1$ and
$\bar\Lambda$.

The HQET parameters appear in the expansion of the heavy meson masses
in terms of the charm and bottom quark masses:
\begin{mathletters}\label{cbmasses}
\begin{eqnarray}
m_B &=&m_b+\bar\Lambda-{\lambda_1+3\lambda_2\over2m_b}+\ldots\,,\\
m_D &=&m_c+\bar\Lambda-{\lambda_1+3\lambda_2\over2m_c}+\ldots\,.
\end{eqnarray}
\end{mathletters}%
For each set of values $\{\bar\Lambda,\lambda_1,\lambda_2\}$,
eqs.(\ref{cbmasses}) determine $m_c$ and $m_b$.
The consistency of the heavy quark expansion requires that these values
of the quark masses are used in the theoretical expressions for the decay
rates.  Then, a precise knowledge of $\bar\Lambda$ and $\lambda_1$ would
allow accurate predictions for heavy hadron decays.

It was noticed in ref.\cite{LS} that semileptonic $D$ decays provide
a lower bound on the parameter $\lambda_1$ at each particular value of
$\bar\Lambda$. At present, there exist three measurements of inclusive
rates that provide bounds on $\bar\Lambda$ and $\lambda_1$:
$D\to X\,e\,\nu$, $B\to X\,e\,\nu$, and $B\to X\,\tau\,\nu$.
In what follows we discuss the implications of each of these
processes in turn.

For $D\to X\,e\,\nu$, following ref. \cite{LS} (with minor
modifications explained below), we compute the {\it rate}
$\Gamma(D\to X\,e\,\nu)$ including order $1/m_c^2$ and order
$\alpha_s$ corrections, and compare it to the experimental value
of $BR\,(D\to X\,e\,\nu)/\tau_D$. We use \cite{PDG}
\begin{equation}
BR\,(D^\pm\to X\,e^\pm\,\nu)=17.2\pm1.9\,\%\,,\qquad
\tau(D^\pm)=1.066\pm0.023\,{\rm ps}\,.
\end{equation}
The theoretical expression for the decay rate can be found in
ref.\cite{LS}. Demanding that the theoretical prediction be compatible
with the experimental result, we constrain the allowed range in the
$\bar\Lambda-\lambda_1$ plane.  This is plotted in fig.1.
The allowed region is the area above the curve marked with (1).
A large uncertainty arises in the calculation because the
${\cal O}(\alpha_s^2)$ corrections are unknown. Effectively,
these corrections set the scale at which the ${\cal O}(\alpha_s)$
corrections should be evaluated. The curve in the figure results from
the rather safe assumption that the scale is not above $m_c$.
If one assumes that the scale is larger than some value, say $m_c/3$,
then an upper bound  on $\lambda_1$ can also be obtained.
Such a bound does not play an essential role in our analysis, so
we do not present it here. The interested reader may
consult ref.\cite{LS}.
In deriving our bound we took a more conservative approach
than ref.\cite{LS}.  We take exactly the same ranges for the various
input parameters (except for $\alpha_s$), but allow each of them to
independently vary within their respective 1$\sigma$ bounds.
To indicate the difference between the bounds derived in these two
different ways, assuming $\bar\Lambda>0.237\,$GeV ref. \cite{LS}
quotes $\lambda_1>-0.5\,{\rm GeV}^2$ while we obtain
$\lambda_1>-0.65\,{\rm GeV}^2$.
Our bound is likely to strengthen once ${\cal O}(\alpha_s^2)$
corrections are calculated or if the experimental lower bound
on the branching ratio increases.

The decay $B\to X\,e\,\nu$ is analyzed in a similar way, namely
we calculate the rate $\Gamma(B\to X\,e\,\nu)$ including order
$1/m_b^2$, $1/m_c^2$ and $\alpha_s$ corrections, and compare it to
the experimental value of $BR\,(B\to X\,e\,\nu)/\tau_B$. We use
for the branching ratio \cite{PDG}
\begin{equation}\label{Btoenuexp}
BR\,(B\to X\,e\,\nu)=10.7\pm0.5\,\%\,,
\end{equation}
where we combined the systematic and statistical errors in quadrature,
and for $\tau_B|V_{cb}|^2$ the most recent model independent
determination from exclusive $B\to D^*\ell\,\nu$ decays \cite{CLEO},
\begin{equation}
\sqrt{\tau_B\over1.49{\rm ps}}\,|V_{cb}|= 0.037\pm0.007\,.
\end{equation}
This procedure yields the upper bound on $\bar\Lambda$ corresponding to
the curve marked with (2) in fig.1.  The effect of ${\cal O}(\alpha_s^2)$
corrections is likely to strengthen our bounds (by lowering the
effective scale for $\alpha_s$) but is expected to be small.
The sensitivity to higher order corrections in the relations
(\ref{cbmasses}) is large.  In deriving this bound we included estimates
of these uncertainties.  If the upper bound on $|V_{cb}|$ becomes lower,
that would make this bound significantly stronger.  For example,
$\sqrt{\tau_B\over1.49{\rm ps}}\,|V_{cb}|\leq 0.040$ would yield the
upper bound given by the solid curve in fig.~1.
Similarly, an improvement in the lower bound on the branching ratio
would strengthen our bound.

The decay $B\to X\,\tau\,\nu$ has been investigated in detail in
ref.\cite{us}, where the theoretical expression for the branching ratio
can be found, including order $\alpha_s$ and $1/m_b^2$ corrections.
This decay rate, normalized to the experimentally measured rate for
$B\to X\,e\,\nu$, is a very reliable way of setting bounds on
$\bar\Lambda$ and $\lambda_1$.  The sensitivity to higher order QCD
corrections is minute, as the correction to the ratio of rates is
much smaller than the correction to each of them separately \cite{us}.
In addition, the ratio is independent of the overall
$m_b^5\,|V_{cb}|^2$ factor.
Taking into account the various uncertainties in the same conservative
approach described in ref.\cite{us}, and using the recent measurement
\cite{Aleph} with its statistical and systematic errors combined in
quadrature
\begin{equation}
BR\,(B\to X\,\tau\,\nu) =2.76\pm0.63\,\%\,,
\end{equation}
we obtain a bound excluding large values of both $\bar\Lambda$ and
$\lambda_1$.  This is the curve marked with (3) in fig.1.  This bound
could become much stronger if the lower bound on the $B$ to $\tau$
branching ratio becomes stronger.  To demonstrate that, we plot
(dashed line in fig.1) the bound that would correspond to $\sigma/2$
of the above range, {\it i.e.} $BR\,(B\to X\,\tau\,\nu)\geq2.44\%$.

Each of the three experimental results that we have discussed
determines an allowed {\it band} in the $\bar\Lambda-\lambda_1$ plane.
We have not presented the upper bound from $D\to X\,e\,\nu$ because of
the theoretical uncertainties discussed above.  As for the lower bounds
from $B\to X\,e\,\nu$ and $B\to X\,\tau\,\nu$, at present
they are too weak to be interesting (excluding only negative
values of $\bar\Lambda$ and large negative values of $\lambda_1$).
Therefore, we do not present them either.

We have also investigated the implications of existing data on the
inclusive rare decays $B\to X_s\,\gamma$ and $B\to X_s\,\ell^+\,\ell^-$.
These processes yield no useful constraints at present.
It is interesting to note, however, that as the lower bound on
$m_t$ will increase and the upper bound on $BR\,(B\to X_s\,\gamma)$
will decrease, the resulting bounds will be almost parallel
(in the $\bar\Lambda-\lambda_1$ plane) to the bound provided by
$D\to X\,e\,\nu$ decays.  For example, if experiments find
$m_t\simeq170\,$GeV and $BR\,(B\to X_s\,\gamma)\leq3\times10^{-4}$,
then we will obtain the upper bound given by the dash-dotted curve in
fig. 1.  This will be important as the theoretical uncertainty in this
process \cite{rare} is much smaller than in $D\to X\,e\,\nu$ decays.

Our results are summarized in fig.1.  We learn that the two parameters
$\bar\Lambda$ and $\lambda_1$ are significantly constrained by
existing experimental data.  Our most important result, an upper bound
on $\bar\Lambda$, can be easily read off from this figure:
\begin{equation}\label{Lbound}
\bar\Lambda < 0.7\,{\rm GeV}\,.
\end{equation}

To set an upper bound on $\lambda_1$, we would need to know the size
of the ${\cal O}(\alpha_s^2)$ corrections to semileptonic $D$ decays.
Assuming that the relevant value of the strong coupling constant
$\alpha_s$ in $D$ decays is $\alpha_s(m_c)$,
$\alpha_s\leq2\alpha_s(m_c)$, $\alpha_s\leq3\alpha_s(m_c)$, we obtain
the bounds $\lambda_1<0.6\,{\rm GeV}^2$, $\lambda_1<0.7\,{\rm GeV}^2$,
$\lambda_1<0.8\,{\rm GeV}^2$, respectively.
The bound $\lambda_1\lesssim0.8\,{\rm GeV}^2$ can also be obtained
by combining the QCD sum rules prediction for $\bar\Lambda$
with the experimental results for $BR\,(B\to X\,\tau\,\nu)$ \cite{us}.

To set a lower bound on $\lambda_1$, we need a lower bound on
$\bar\Lambda$. We should mention a theoretical model-independent bound
\cite{GM} $\bar\Lambda>0.237\,$GeV (plotted with a dotted line
in fig. 1). However, in ref. \cite{BU} it is argued that this
bound is not valid. If $\bar\Lambda>0.237\,$GeV, we obtain
$\lambda_1>-0.65\,{\rm GeV}^2$ (and also $\lambda_1<1.2\,{\rm GeV}^2$).
The QCD sum rule prediction $\bar\Lambda\geq0.50\,$GeV
\cite{review,BBBD} yields $\lambda_1\geq-0.10\,{\rm GeV}^2$.

It is instructive to compare these bounds to various model calculations.
Our upper bound on $\bar\Lambda$ is hardly weaker than the allowed
range of this parameter predicted by QCD sum rules,
$\bar\Lambda=0.57\pm0.07\,$GeV \cite{review,BBBD}.
The QCD sum rules estimates of the parameter $\lambda_1$ varied
strongly over the past few years \cite{SRlam}. These sum rules are
rather unstable, thus their predictions are uncertain and have large
errors.  Recent progress in this direction \cite{virial},
however, indicates that $\lambda_1$ is likely to be negative and small in
magnitude, $-0.3\lesssim\lambda_1<0\,{\rm GeV}^2$ \cite{Neubert}.
This is certainly allowed by our bounds, but then $\bar\Lambda$ is
likely to be smaller than $0.5\,$GeV.
A recent calculation of $\lambda_1$ in the ACCMM model results
$\lambda_1\simeq-0.08\,{\rm GeV}^2$ \cite{shape}.

These experimental constraints on $\bar\Lambda$ and $\lambda_1$ can be
easily translated into bounds on the charm and bottom quark masses.
We find that in the experimentally allowed region
\begin{equation}\label{masses}
m_c\geq1.43\,{\rm GeV}\,, \qquad m_b\geq4.66\,{\rm GeV}\,.
\end{equation}
We do not quote upper bounds on the quark masses, both because of the
uncertain status of the model independent lower bound on $\bar\Lambda$,
and because such a bound would also depend on assumptions about the
higher order QCD corrections to semileptonic $D$ decays.

A precise knowledge of the bottom and charm quark masses (and of
$\lambda_1$) is important, for example, for a model independent
determination of $|V_{cb}|$ from inclusive semileptonic
$B\to X_c\,\ell\,\nu$ decays.
When phenomenological constraints, like those derived in this paper,
or possible theoretical model-independent bounds on $\bar\Lambda$ and
$\lambda_1$ will become more restrictive, inclusive $B\to X_c\,\ell\,\nu$
decays may provide an alternative determination of $|V_{cb}|$,
comparable in accuracy to that from the zero recoil limit of exclusive
$B\to D^{(*)}\ell\,\nu$ decays.

The bounds for the quark masses in eq. (\ref{masses}) have important
consequences for $BR\,(B\to X\,e\,\nu)$.
The theoretical result for the semileptonic {\it branching ratio}
derived by calculating both the semileptonic and hadronic widths,
seems to constitute a puzzle at present \cite{BBSV}: theory predicts
$BR\,(B\to X\,e\,\nu)>0.12$, in conflict with the experimental result
(\ref{Btoenuexp}).  While we have nothing to add here to the ongoing
discussion on the uncertainties in the calculation of the hadronic
width, we would like to recall that a solution of the problem prefers
low quark masses \cite{AP}.
Thus, our lower bounds on $m_c$ and $m_b$ point into the direction
of increasing this problem, rather than eliminating it.

To summarize our main results, we showed that a combination of
experimental data on semileptonic heavy meson decays gives the
upper bound $\bar\Lambda<0.7\,$GeV and lower bounds on the heavy
quark masses: $m_c\geq1.43\,$GeV and $m_b\geq4.66\,$GeV.
Improved experimental results or theoretical calculations of
${\cal O}(\alpha_s^2)$ corrections will make these constraints
significantly stronger.

\acknowledgments
It is a pleasure to thank Adam Falk, Yuval Grossman, Matthias Neubert
and Adam Schwimmer for useful discussions.

\begin{figure}
\caption[aa]{
The allowed range in the $\bar\Lambda-\lambda_1$ plane.
The bounds are depicted by solid lines shaded on their excluded sides:
(1) from $D\to X\,e\,\nu$ decays; (2) from $B\to X\,e\,\nu$ decays;
(3) from $B\to X\,\tau\,\nu$ decays.
The solid line would be the upper bound if experiment finds
$\protect\sqrt{\tau_B\over1.49{\rm ps}}\,|V_{cb}|\leq 0.040$.
The dashed line would be the upper bound if experiment finds
$BR\,(B\to X\,\tau\,\nu)\geq2.44\%$.
The dash-dotted curve would be the upper bound if experiment finds
$BR\,(B\to X_s\,\gamma)\leq3\times10^{-4}$ and $m_t\simeq170\,$GeV.
}
\end{figure}

\end{document}